\documentstyle[aps,twocolumn,epsfig,tighten,floats]{revtex}
\newcommand{\txl}{T$\chi$L~}
\newcommand{\tr}{\mbox{tr}}
\def\Journal#1#2#3#4{{#1} {\bf #2}, #3 (#4)}

\def\ea{$\mbox{\it et al }$}

\def\NPB{{\em Nucl. Phys.} B}
\def\NPA{{\em Nucl. Phys. Proc. Suppl.} A}
\def\NPP{{\em Nucl. Phys. Proc. Suppl.} B}
\def\PLB{{\em Phys. Lett.}  B}
\def\PRL{\em Phys. Rev. Lett.}
\def\PRD{{\em Phys. Rev.} D}

\begin{document}
\draft
\title
{\hfill
\begin{minipage}{0pt}\scriptsize \begin{tabbing}
\hspace*{\fill} HLRZ2000-14\\ 
\hspace*{\fill} GUTPA/00/09/01\\
\hspace*{\fill} WUP-TH 00-17 
\end{tabbing} 
\end{minipage}\\[8pt] 
Flavour singlet pseudoscalar masses\\
in $N_f = 2$ QCD} \author{T. Struckmann$^a$,
K.~  Schilling$^{a,b}$\\
  G.~Bali$^c$, N.~Eicker$^b$ ,  S.~G\"usken$^b$, Th.~Lippert$^b$,\\
  H.~Neff$^a$, B.~Orth$^b$, W.~Schroers$^b$, J.~Viehoff$^a$, and
  P.~Ueberholz$^b$ }
\address{${}^{a}$NIC, Forschungszentrum J\"ulich, 52425 J\"ulich and\\
  DESY, 22603 Hamburg, Germany}  
\address{${}^{b}$Fachbereich
  Physik, Bergische Universit\"at, Gesamthochschule
  Wuppertal\\ Gau\ss{}stra\ss{}e 20, 42097 Wuppertal, Germany}
\address{${}^{c}$Department of Physics and Astronomy, 
The University of Glasgow,
Glasgow G12 8QQ, Scotland\\[.2 cm]
  (SESAM - T$\chi$L Collaboration)} 
%\date{\today}
\maketitle
\begin{abstract}
  We perform a lattice mass analysis in the flavour singlet pseudoscalar
  channel on the SESAM and \txl full QCD vacuum configurations, with 2 active
  flavours of dynamical Wilson fermions at $\beta = 5.6$.  At our inverse
  lattice spacing, $a^{-1}\approx 2.3$~GeV, we retrieve by a chiral
  extrapolation to the physical light quark masses the value $m_{\eta'} =
  3.7_{-4}^{+8} m_{\pi}$.  A crude extrapolation from ($N_f=3$) phenomenology
  would suggest $m_{\eta'}\approx 5.1 m_{\pi}$ for $N_f=2$ QCD.  We verify
  that the mass gap between the singlet state $\eta'$ and the $\pi$ flavour
  triplet state is due to gauge configurations with non-trivial topology.
%\pacs{PACS numbers: 11.15.Ha, 12.38.Gc, 12.39.Mk, 12.39.Pn}
\end{abstract}
\narrowtext
\section{Introduction}
Lattice gauge theory (LGT) has been established as the standard method to deal
with infrared aspects of quantum chromodynamics (QCD).  Recently,
the light hadronic flavour non-singlet masses have been accurately determined
by $N_f=2$ QCD simulations on the teracomputing scale~\cite{cppacs}.
Unfortunately, the situation is much less clear when it comes to the
interesting physics of flavour symmetric hadronic states (like the
$\eta'$-meson) which are expected to be influenced by the topological
properties of the QCD vacuum: we have to await multi-teracomputing to see it
settled. 

The problem is due to the very occurrence of Zweig-rule forbidden
contributions to the singlet hadronic propagator in form of disconnected
diagrams, as already known from early  feasibility studies in
quenched QCD~\cite{kuramashi}.  The difficulties arise for three reasons: (a)
{\it eo ipso}, such disconnected correlators induce a high level of gauge field
noise into the calculations; (b) their computation is costly as it involves
momentum zero projections of quark loops, the evaluation of which requires the
use of stochastic estimator techniques; (c) the propagator of a flavour
singlet pseudoscalar meson, $C_{\eta'}$, turns out to be the {\it difference}
between connected  and disconnected  diagrams with
the possibility of numerical cancellations.

Indeed, at large Euclidean time separations, $t$, these cancellations are
doomed to be strong  if they are to render the large empirical flavour
singlet/non-singlet mass gap~\footnote{In our $N_f=2$ world we have a triplet
  (rather than an octet) of flavour non-singlet mesons.  Moreover, working
  with mass-degenerate quarks, our $\pi$'s are exactly mass degenerate too.},
 $M_0^2 \equiv  M_{\eta'}^2 - M_{\pi}^2$.  As a consequence, however, the
signal-to-noise ratio becomes a serious problem for the direct lattice
approach to flavour singlet objects and makes it hard to keep control on
systematic errors.  For all these reasons, {\it ab initio} full QCD lattice
investigations of the $\eta'$ mass have not yet overcome an exploratory stage.
There is of course a way to avoid all this by following an indirect strategy
and taking resort to  the assumptions underlying the Witten-Veneziano
formula~\cite{witten_veneziano}; this workaround amounts to determining the
mass gap from {\it quenched} lattice determinations of the topological
susceptibility~\cite{pisa}.

Previous pioneering work in {\it full} QCD largely focussed on a two-step
recipe to deal with the above problems: {\it (i)} determine $m_{\pi}$ at large
$t$ and {\it (ii)} compute the mass gap~\footnote{Upper (lower) case letters
  refer to masses in physical (lattice) units.}, $m_0$, from the ratio of
connected and disconnected correlators, $R(t) =
C_{disc}(t)/C_{conn}(t)$~\cite{kilcup,AliKhan:1999zi,Michael:1999rs} in the range
of smallish $t$-values.  In our present approach we seek for a $t$-window
within which a straightforward (one-step) flavour singlet propagator analysis
can be pertinently achieved.

In a recent study we already applied improved stochastic estimator techniques
--- as geared previously for coping  with disconnected operator insertions in
the context of hadronic matrix elements~\cite{viehoff:stochastic} --- on the
flavour singlet correlators, with pointlike sources~\cite{osaka,struckmann:thesis}.
In this paper we shall show by a mass plateau analysis that a standard mass
computation on the flavour singlet propagator itself will become feasible with
reasonable control of systematic errors, once smeared operators are used.
 
\section{Lattice prerequisites}
\label{section:observables}
We consider the pseudoscalar flavour singlet operator in a flavour symmetric
theory
\begin{equation}
S(x) =  \sum_{i=1}^{N_f} \overline{q}_i(x) \gamma_5 q_i(x) \ ,
\label{eq:operator}
\end{equation}
with $N_f$ flavours. By the usual  Wick contraction it leads
 to the flavour singlet propagator in terms of the inverse Dirac operator,
$\Delta \equiv D^{-1}$:
\begin{eqnarray}
\label{eq:greens}
 C_{\eta'}(0|x) & \sim & \langle N_f \tr(\Delta(0|x)\Delta^{\dagger}(0|x)) \nonumber \\
 && - N_f^2 \tr(\gamma_5\Delta^{\dagger}(0|0)) \tr(\gamma_5\Delta(x|x))\rangle \; ,
\end{eqnarray}
%\begin{equation}
%\label{eq:greens}
% C_{\eta'}(0|x) \sim N_f \tr(\Delta(0|x)\Delta^{\dagger}(0|x)) 
% - N_f^2 \tr(\gamma_5\Delta^{\dagger}(0|0)) \tr(\gamma_5\Delta(x|x)) \; ,
%\end{equation}
which is a sum of fermionic connected and disconnected contributions with
traces to be taken in the spin and colour spaces. In the rest of the paper we
shall refer to them as `one-loop' and `two-loop' contributions, respectively.
The traces are computed with $Z_2$ noise sources including diagonal
improvement as explained in ref.~\cite{osaka}.

The momentum zero projection
\begin{equation}
C_{\eta'}(t) \equiv \langle S(t)S(0)\rangle_{conn} - \langle S(t)S(0)\rangle_{disc} 
\label{eq:gvont}
\end{equation}
is expected to decay exponentially, 
$\sim \exp(-m_{\eta'}t)$, and thus to reveal the flavour singlet mass, $m_{\eta'}$.
On a toroidal lattice with temporal extent $T$ one should encounter the usual
$\cosh$ behaviour at large values of $t$ and $T-t$
\begin{equation}
C_{\eta'}(t) \rightarrow \exp(-m_{\eta'}t) + \exp(-m_{\eta'}(T-t))\; .
\end{equation}
From this parametrization effective masses, $m_{\eta'}^t$, can be retrieved
by solving the implicit equations
\begin{equation}
\frac{C_{\eta'}(t+1)}{C_{\eta'}(t)} = 
\frac{\exp(-m_{\eta'}^t(t+1))+\exp(-m^t_{\eta'}(T-t-1))}{\exp(-m^t_{\eta'}t) +
 \exp(-m^t_{\eta'}(T-t))}\; .
\label{eq:effmasses}
\end{equation}

For sufficiently large values of $t$, the effective masses should saturate
into a plateau.  The crucial question is, however, whether one can establish a
{\it $t$-window of observation} that reveals a definite plateau behaviour of
$m^t_{\eta'}$ before noise takes over.

\subsection{Operator  smearing}
The building blocks for hadronic observables are the quark propagators, $\xi$,
which may be computed by solving the (discretized) Dirac equation  with
appropriate source vectors, $\phi_s(z)$, on the lattice:
\begin{equation}
D(z,x)\xi(x)  = \phi_s(z) \; .
\label{eq:dirac}
\end{equation}
In standard spectrum analysis one generally applies some kind of spatial
smearing to the hadron source (located at $t=0$) in order to enhance the
ground state signals of the resulting hadronic operators at medium values of
$t$.  Needless to say, this appears to be all the more necessary in the
present context where -- as explained above -- we are faced both with (a)
cancellations (between $C_{conn}$ and $C_{disc}$) and (b) noisier signals
(from $C_{disc}$).

We used our smearing procedure as applied in the analysis of light non-singlet
masses~\cite{sesam:masses}; it is characterized by $N$ diffusive iteration  steps
\begin{equation}
\label{eq:smearing}
\phi_s^{(i+1)}(x) = \frac{1}{1+6\alpha}\left[\phi_s^{(i)}(x,t) +\alpha\sum_{\mu}
 \phi_s^{(i)}(x + \mu )^{\mbox{{\tiny  p.t.}}}\right] \; ,
\end{equation}
where the index `p.t.' stands for `parallel transported', and the sum extends
over the six spatial neighbours of $x$. For $\phi_s^{(0)}$ we start out from
pointlike sources for the connected and $Z_2$-noise nonlocal sources for the
disconnected diagrams. In this way the bilinear quark operators,
Eq.~(\ref{eq:operator}), were computed after $N=25$ such smearing steps, with
the value $\alpha = 4.0$.  The smearing procedure was applied to meson sources
as well as to sinks -- both for $C_{disc}$ and $C_{conn}$, in order to correctly
maintain their relative normalisations.

\begin{table} 
\caption{Simulation parameters used at $\beta = 5.6$ and  numbers of
stochastic sources (used with local and smeared operators, $N_{est}^{ll}$,
$N_{est}^{sm}$). Last column: numbers of available decorrelated 
vacuum field configurations, $N_{conf}$.}
\label{tab:simdat} 
\begin{tabular}{|cccccc|}
\hline
$\kappa_{sea}$ & $m_{\pi}/m_{\rho}$ & $L^3*T$ & $N_{est}^{ll}$ & $N_{est}^{sm}$
& $N_{conf}$ \\ 
\hline 0.1560 & 0.834(3)  & $16^3*32$ & $400$ & $400$ & $195$ \\
\hline
0.1565 & 0.813(9) & $16^3*32$ & $400$ & $400$ & $195$ \\
\hline
0.1570 & 0.763(6) & $16^3*32$ & $400$ & $400$ & $195$ \\
\hline
0.1575 & 0.692(10) & $16^3*32$ & $400$ & $400$ & $195$ \\
\hline
0.1575 & 0.704(5)  & $24^3*40$ & $400$ & $100$ & $156$ \\
\hline
0.1580 & 0.574(13) & $24^3*40$ & $100$ & $100$ & $156$ \\
\hline
\end{tabular}
\end{table}

In table~\ref{tab:simdat} we list the run parameters of our simulations, which
make use of vacuum field configurations generated by the SESAM ($16^3\times
32$ lattice~\cite{sesam}) and the T$\chi$L ($24^3\times 40$
lattice~\cite{txl}) collaborations, both with $N_f = 2$ and $\beta = 5.6$. We
have used five different sea quark masses and two different lattice sizes to gain
some control on finite-size effects. While the number of vacuum configurations
varies from $156$ to $195$, the number of independent stochastic sources has
been chosen to be $400$ on the small lattices, both for local ({\it ll}) and
smeared ({\it sm}) operators. On the large lattices $100$ ($400$ for
$\kappa_{sea}=0.1575$-{\it ll}) source vectors were used.

\begin{figure}[!htb]
\centering{\epsfig{figure=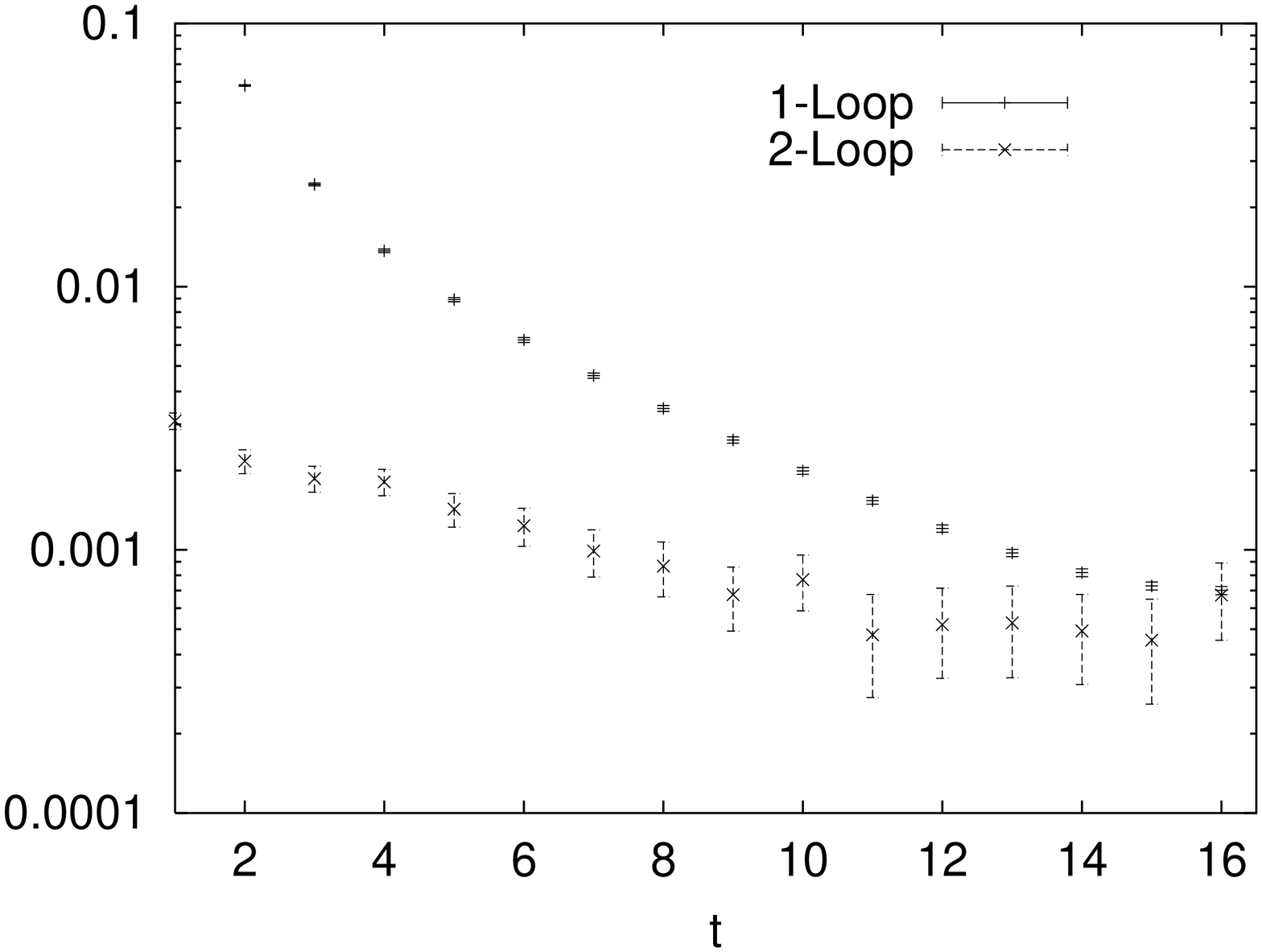,angle=270,width=\columnwidth}} 
\centering{\epsfig{figure=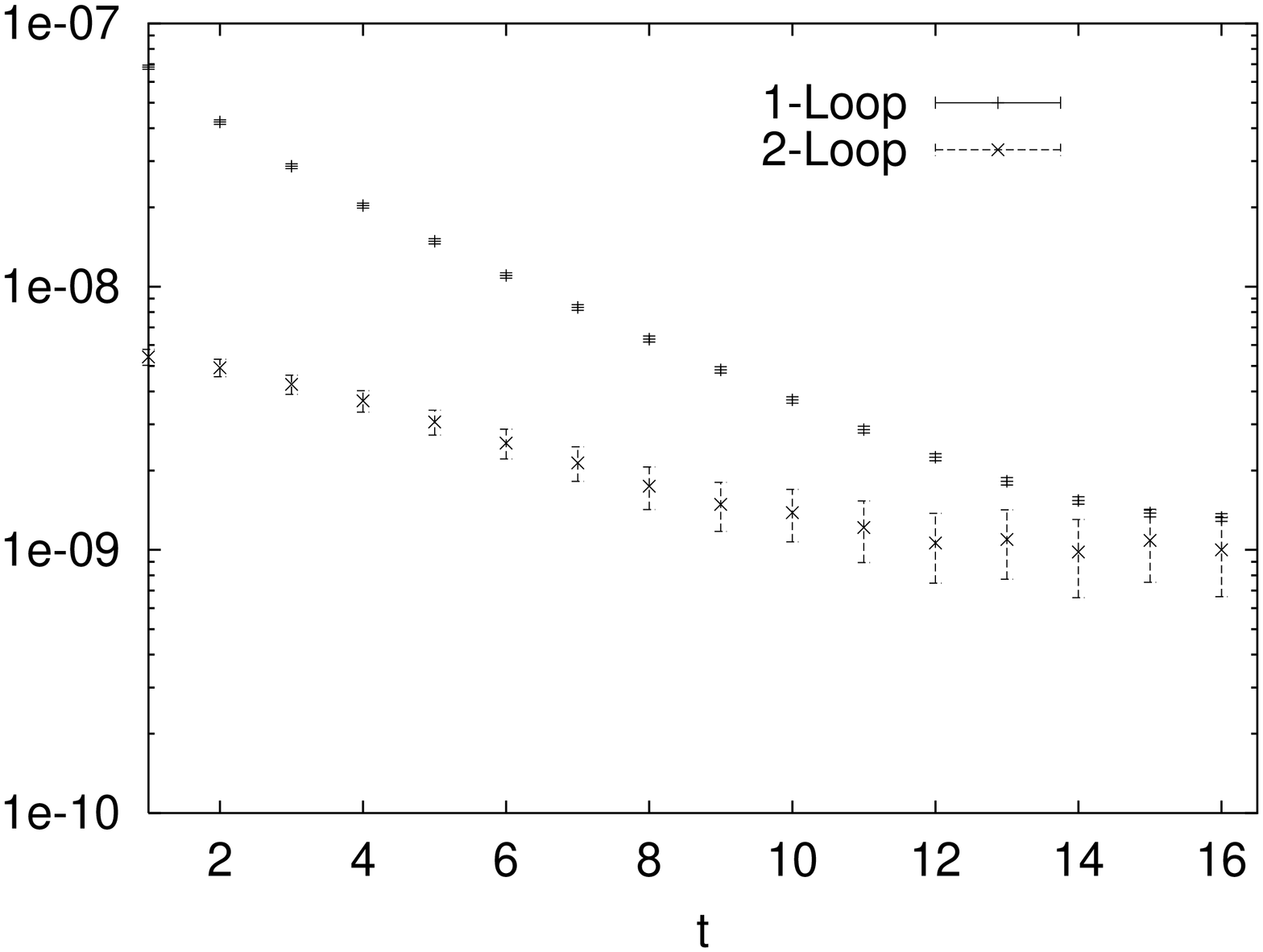,angle=270,width=\columnwidth}}

\vskip .4 cm
\caption{Effect of smearing on the correlation functions.
  SESAM configurations  at the lightest sea quark mass, $\kappa_{sea} =
  0.1575$. Top figure:  local operator; bottom figure: smeared operator.
  Upper (lower) data set refers to one-loop (two-loop) contributions.}
\label{fig:correlators}
\end{figure}
Fig.~\ref{fig:correlators} illustrates the quality of our data in terms of the
one-loop and two-loop correlators at the lightest sea quark
mass on the SESAM lattice, computed with pointlike (upper
figure) and smeared operators (lower figure), using $N_{est} = 400$
stochastic $Z_2$-noise sources with diagonal 
improvement~\cite{osaka}.  The errors quoted are statistical and have been
obtained by jackknifing.  We find a marked improvement of the signal 
by help of source smearing in the regime $5 \leq t \leq 12$.

\begin{figure}[!htb]
\centering{\epsfig{figure=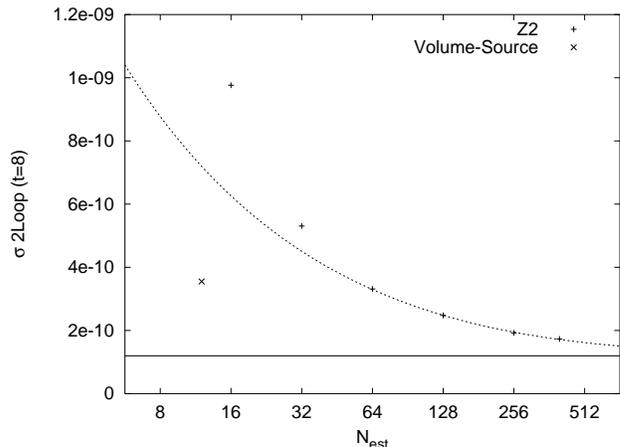,angle=270,width=\columnwidth}}
\caption{Error, $\sigma$, of two-loop 
  signals versus $N_{est}$ at $\kappa_{sea} = 0.1570$ on
  time slice $t=8$, for smeared sources and sinks. The curve is the best fit
  according to the parametrisation given by Eq.~(\ref{eq:staterr}).}
\label{fig:error_estimate} 
\end{figure}

One should remember that the data points in Fig.~\ref{fig:correlators} suffer
from two kinds of stochastic noise: the one from the gauge fields and the one
from the noisy sources.  In order to disentangle them it is highly instructive
to study the error on the two-loop signal, $\sigma$, as a function of the
number of stochastic sources, $N_{est}$ . In Fig.~\ref{fig:error_estimate} we
plot this quantity for the SESAM ensemble at $\kappa_{sea} = 0.1570$ on a
time slice of interest, $t = 8$.  At sufficiently large values of $N_{est}$
the parametrization
\begin{equation}
%\sigma^2 = \frac{S_{est}^2}{N_{conf}*N_{e}} +
%\frac{S_{conf}^2}{N_{\mbox{conf}}}
\sigma^2 = \frac{\Sigma_{est}^2}{N_{est}} + \Sigma_{conf}^2 \; 
\label{eq:staterr}
\end{equation}
is expected to describe the superposition of errors from source and gauge
field fluctuations. Therefore, the error analysis can provide a useful check
on the quality of the stochastic estimator outputs.  The data in
Fig.~\ref{fig:error_estimate} indeed yield {\rm convincing evidence for early
  asymptotic} $N_{est}$-dependence, with a threshold value $N_{est} \simeq 64$.
Moreover, we find that on the SESAM sample the genuine gauge field noise (as
indicated by the horizontal asymptotic $N_{est} \rightarrow \infty$ line)
prevails once we choose $N_{est} \geq 100$. As to the subasymptotic regime,
 one might attribute the apparent `non-standard' behaviour of
$\sigma$ to pollutions from subleading, non-trace terms in the
stochastic estimate of the loops.

The main message from Fig.~\ref{fig:error_estimate} is that the $Z_2$ noise
method does provide reasonable parametric control over the additional
fluctuations induced by the stochastic estimator on the observable.
This control cannot be taken for granted when applying the volume source
technique ($12$ color-spin explicit inversions per configuration) which
refrains from using stochastic sources and relies fully on gauge invariance
and gauge noise for the suppression of nondiagonal contributions to the trace
estimates~\cite{kuramashi}.  For comparison, however, we have included the
corresponding error as we computed  it on our gauge field ensemble.
\begin{figure}[!htb]
\centering{\epsfig{figure=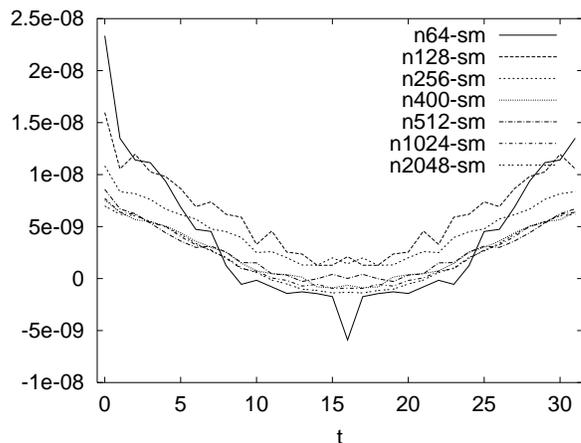,angle=270,width=\columnwidth}}
\caption{t-dependence of the two-loop signal,
 on a single configuration with smeared operator,  for
  various values of $N_{est}$.  }
\label{fig:large_est} 
\end{figure}

Complementary to these considerations one may  study the overall (in $t$)
effects of finite source sampling on the estimate of the two-loop correlator
for  a particular gauge configuration.  Fig.~\ref{fig:large_est} illustrates,
again at $\kappa_{sea} = 0.1570$ on the small lattice, the kind of fluctuations
induced by the stochastic sources of the two-loop correlator with smeared
operators at various
\begin{figure}[t]
\begin{minipage}[b]{.92\linewidth}
\centering{\epsfig{figure=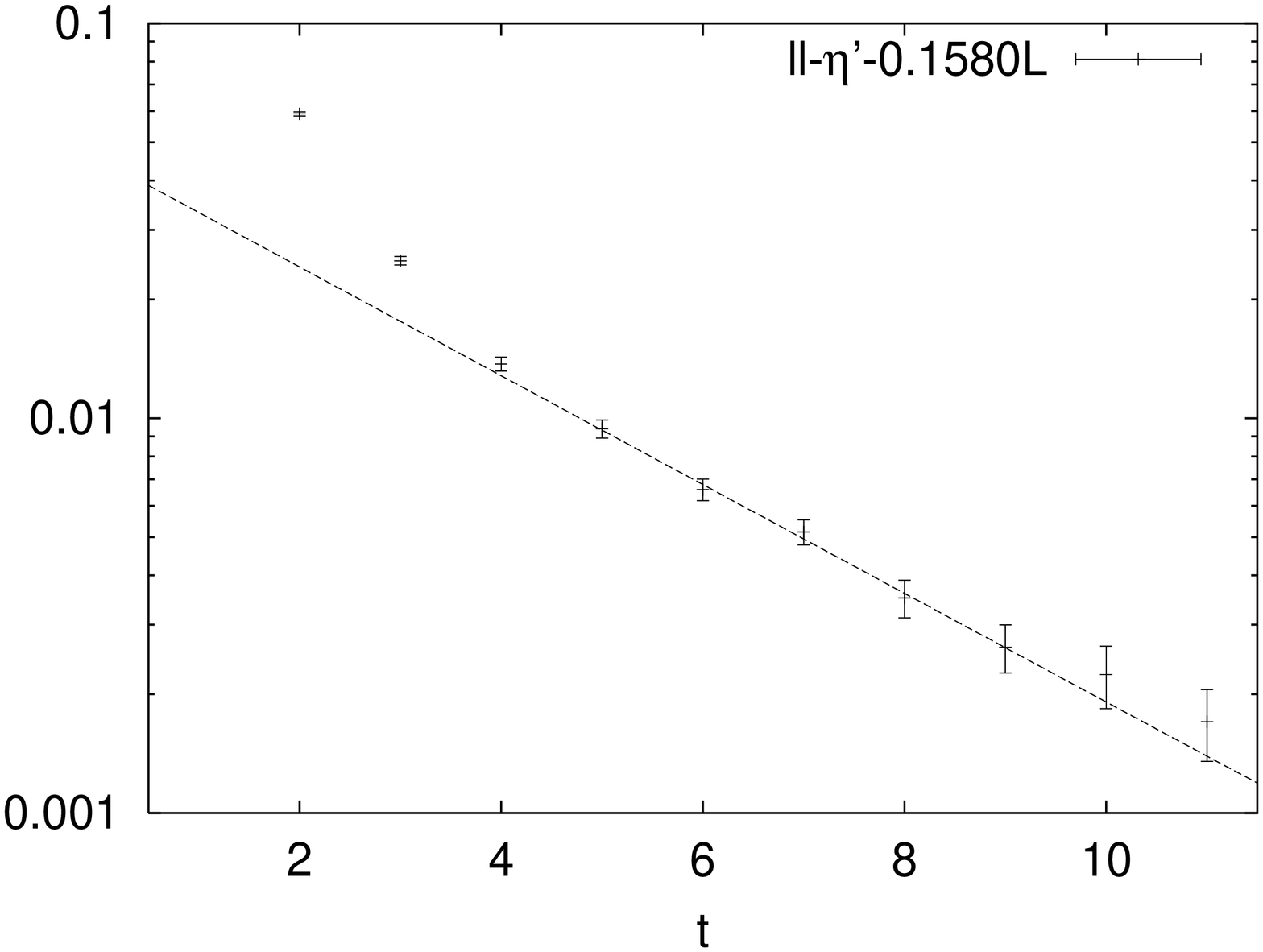,angle=270,width=1.1\linewidth}}
\end{minipage} \\
\begin{minipage}[b]{.92\linewidth}
\centering{\epsfig{figure=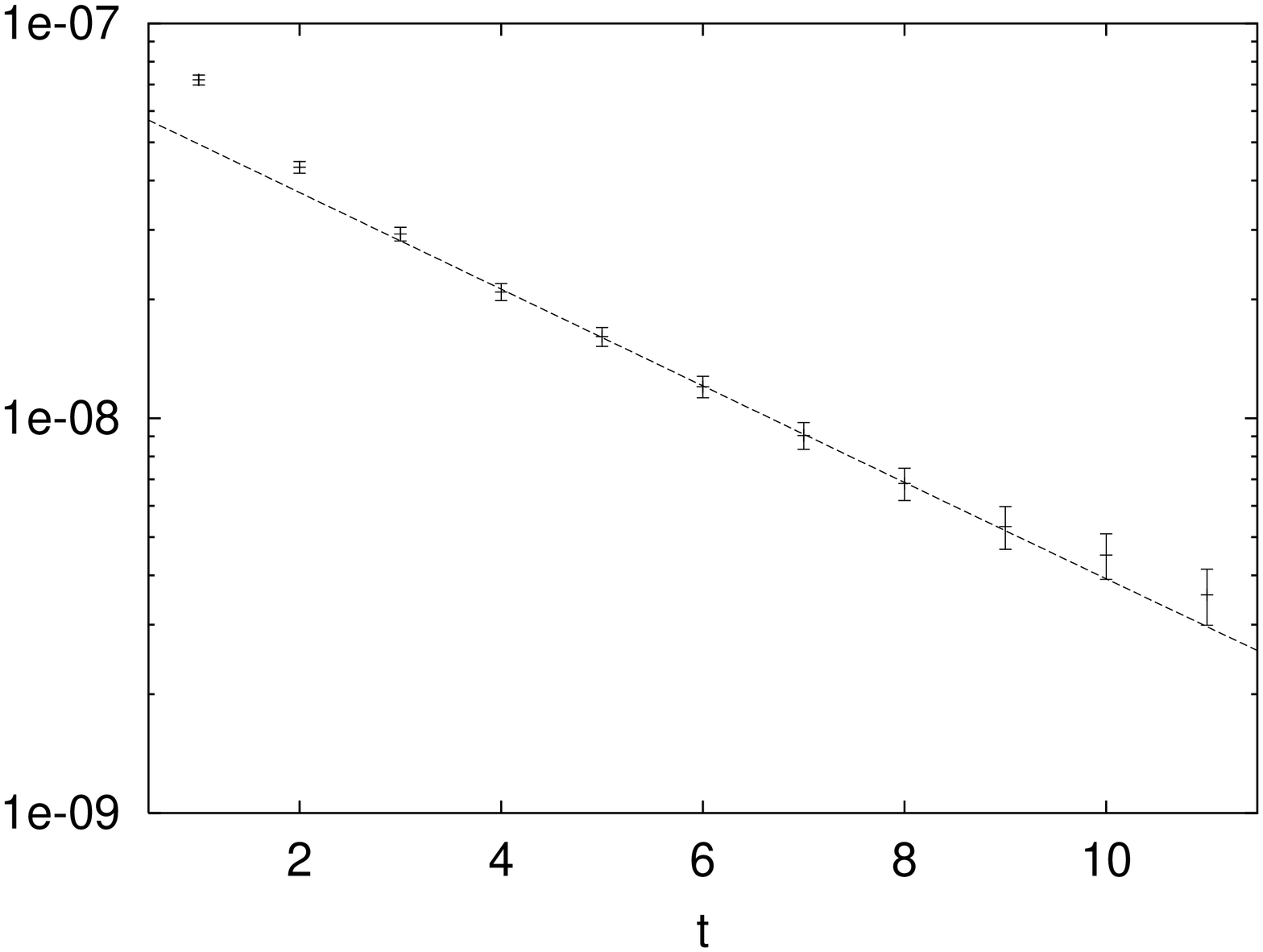,angle=270,width=1.1\linewidth}}
\end{minipage} \hfill
\caption{Ground state dominance of the $\eta'$-propagator ($\kappa_{sea} = 0.1580$), with smeared 
  sources and sinks. Top: with local operators, bottom: with smeared operators.}
\label{2lfit}
\end{figure}
values of $N_{est}$ which we ran up to $2048$ in this case. It appears that on
our sample the $Z_2$ noise injected from the sources into the correlator is
adequately suppressed at $N_{est} \approx 400$. This justifies again that on
the small lattices $N_{est} =400$ is a reasonable choice for the present study.  On
the large lattices, however, enhanced self-averaging effects allow  for a
smaller number of stochastic sources,  $N_{est} \simeq 100$.
\begin{table}[b]
\begin{center}
\caption{Fit ranges\label{fittab}} 
\hspace{0.2cm}
\begin{tabular}{|cccccc|}
\hline
$\kappa_{sea}$ & $L^3*T$ & $\pi$-ll-Fit & $\eta'$-ll-Fit & $\pi$-sm-Fit & $\eta'$-sm-Fit \\
\hline
0.1560 & $16^3*32$ & $12-16$      & $6-9$          & $9-16$      & $5-10$          \\
\hline
0.1565 & $16^3*32$ & $13-16$      & $6-9$          & $9-16$      & $5-10$          \\
\hline
0.1570 & $16^3*32$ & $12-15$      & $6-9$          & $9-16$      & $5-10$          \\
\hline  
0.1575 & $16^3*32$ & $12-15$      & $6-9$          & $9-16$      & $6-11$          \\
\hline    
0.1575 & $24^3*40$ & $12-15$      & $6-9$        & $9-16$      & $5-9$          \\
\hline   
0.1580 & $24^3*40$ & $12-15$      & $6-9$          & $9-16$      & $5-9$          \\
\hline
\end{tabular}
\end{center}
\end{table}

We are now in the position to study plateau formation on our ensemble of
vacuum configurations.

\subsection{Plateaus of effective masses from smearing}

The effect of smearing on the flavour symmetric correlator, $C_{\eta'}$, is
visualized in the comparative twin plot of  Fig.~\ref{2lfit}, as
obtained at our smallest quark mass on the \txl lattice.  {\it Prima vista} we
do find reasonable signals on this correlator up to $t \approx 10$.  Moreover,
by inspection of the $\cosh$-fits, we find a considerable decrease of excited
state contributions as a result of smearing.
\begin{figure}[!htb]
\centering{\epsfig{figure=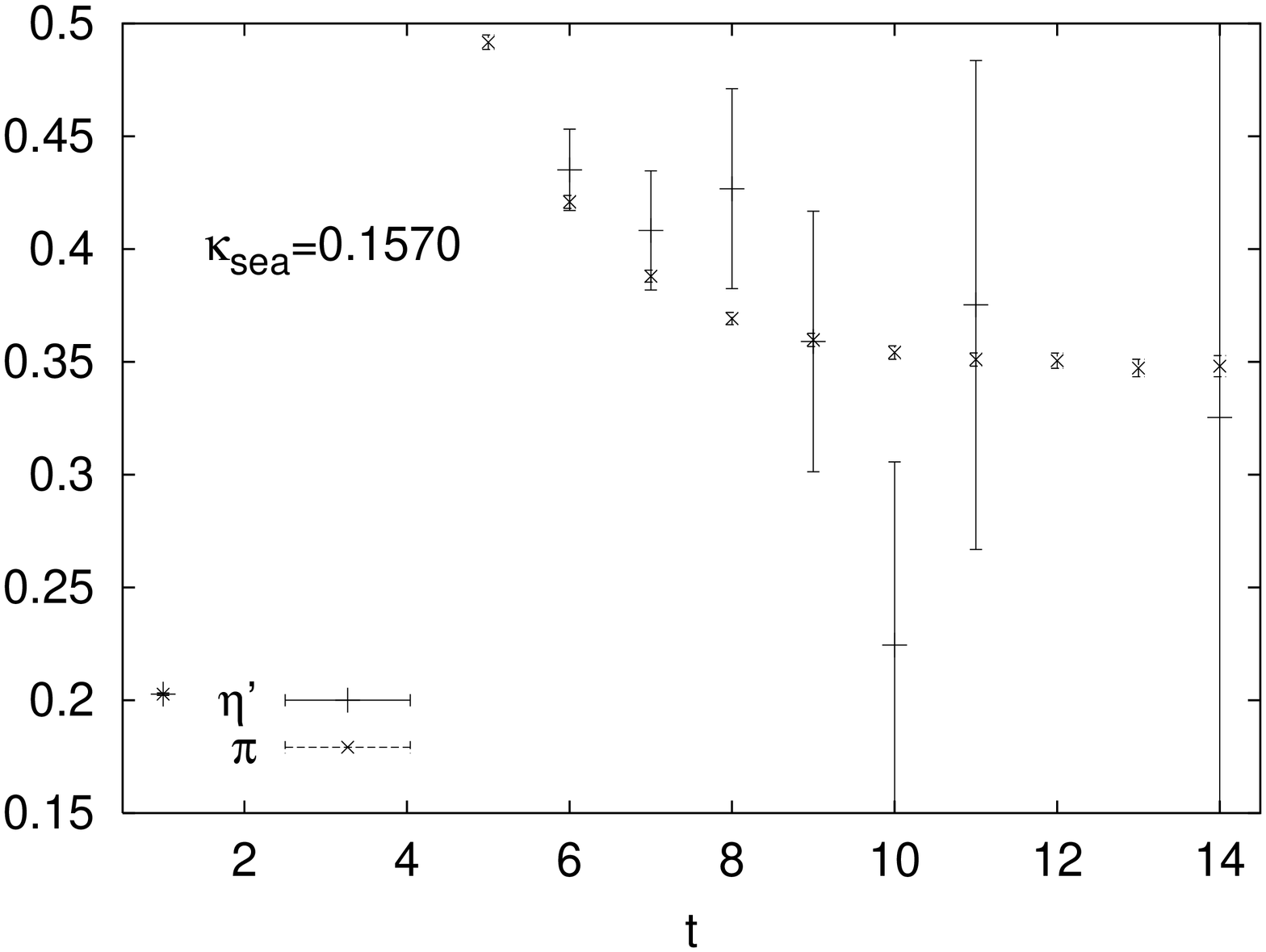,angle=270,width=\columnwidth}}
\centering{\epsfig{figure=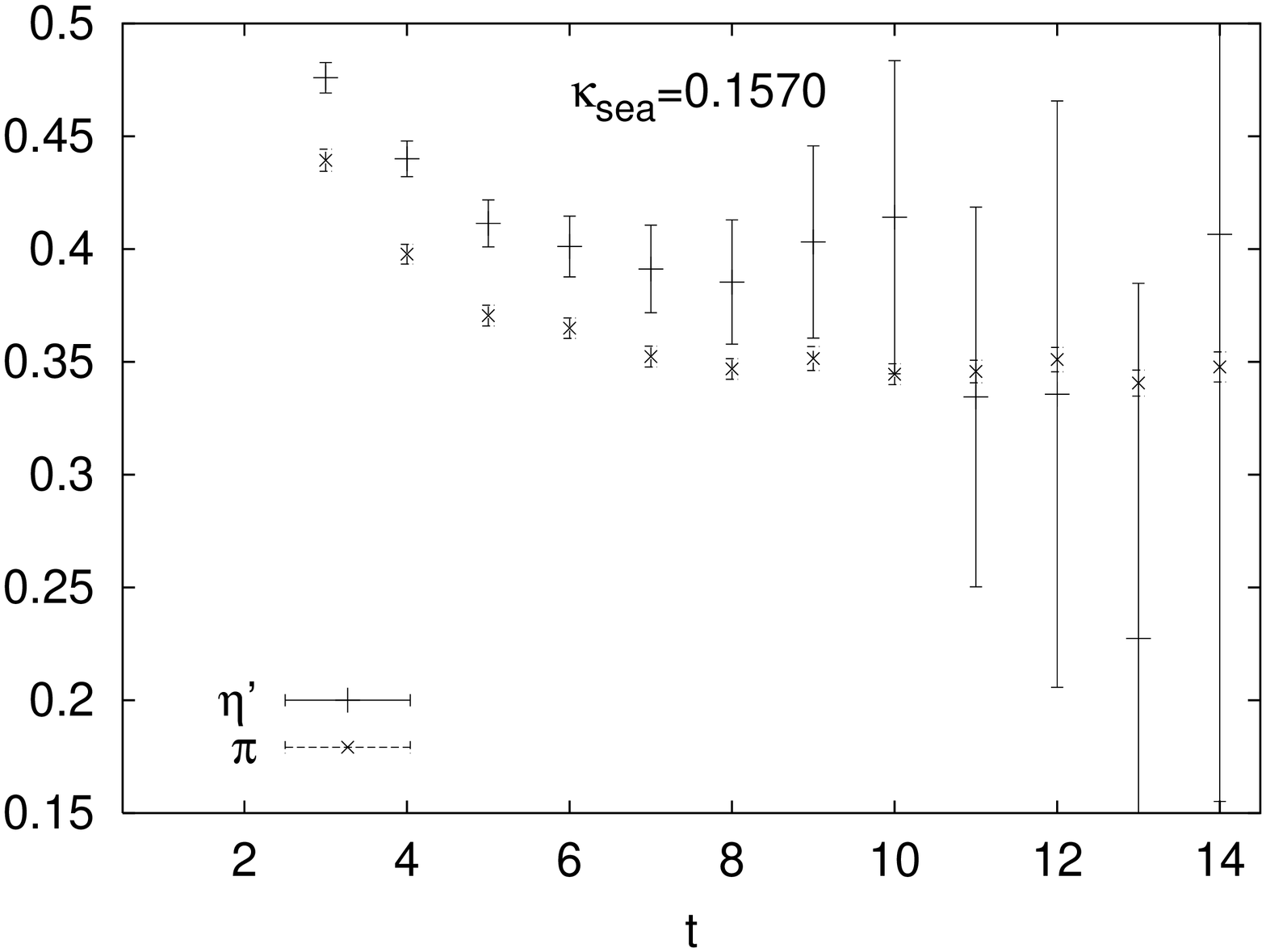,angle=270,width=\columnwidth}}
\caption{Effective masses  in the flavour singlet and `octet' channels, with pointlike (top) and 
  smeared operators (bottom), at $\kappa_{sea} = 0.1570$.}
\label{fig:eff_loc_smear}
\end{figure}

\begin{figure}[!htb]
\centering{\epsfig{figure=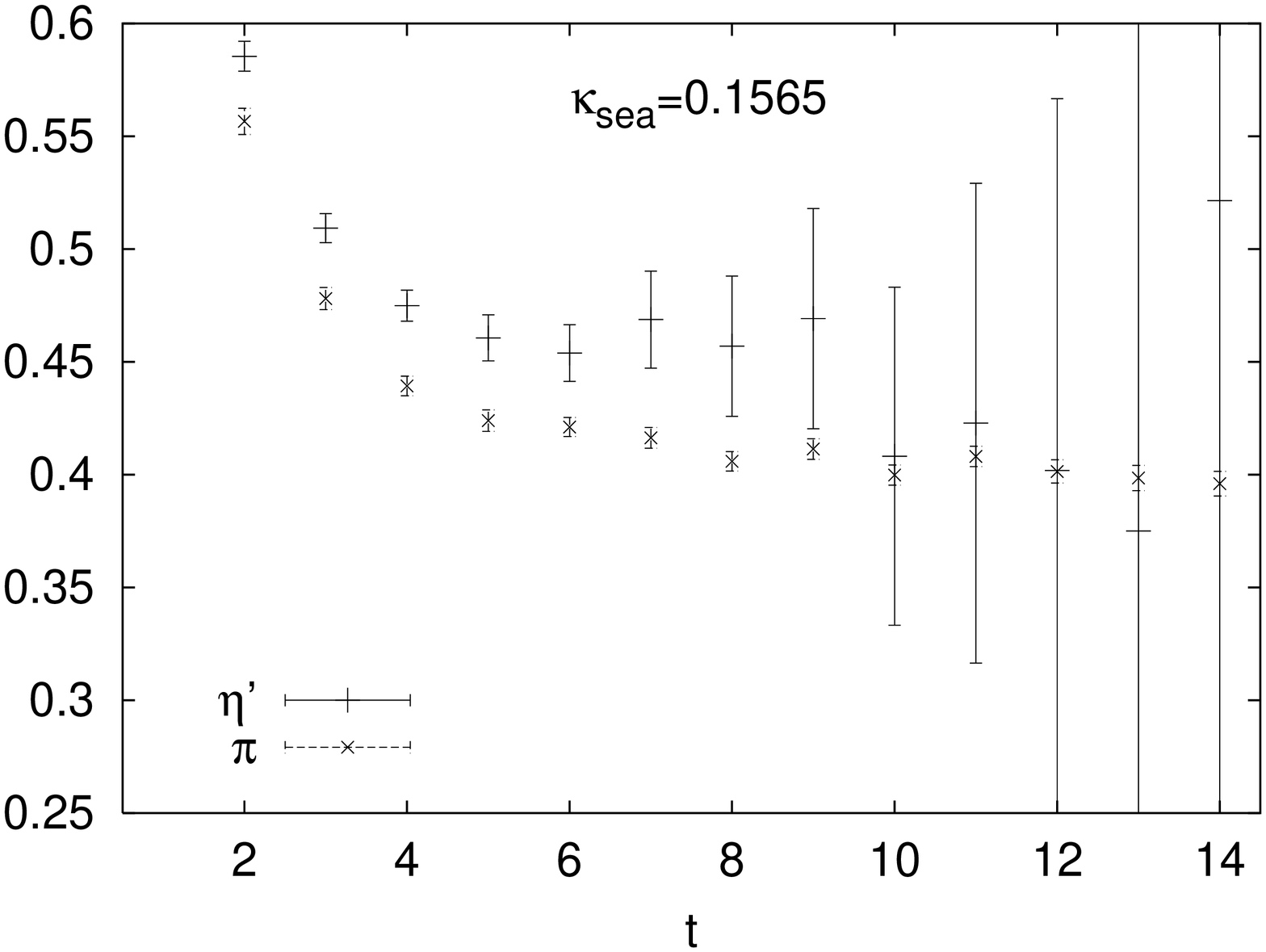,angle=270,width=\columnwidth}}
\centering{\epsfig{figure=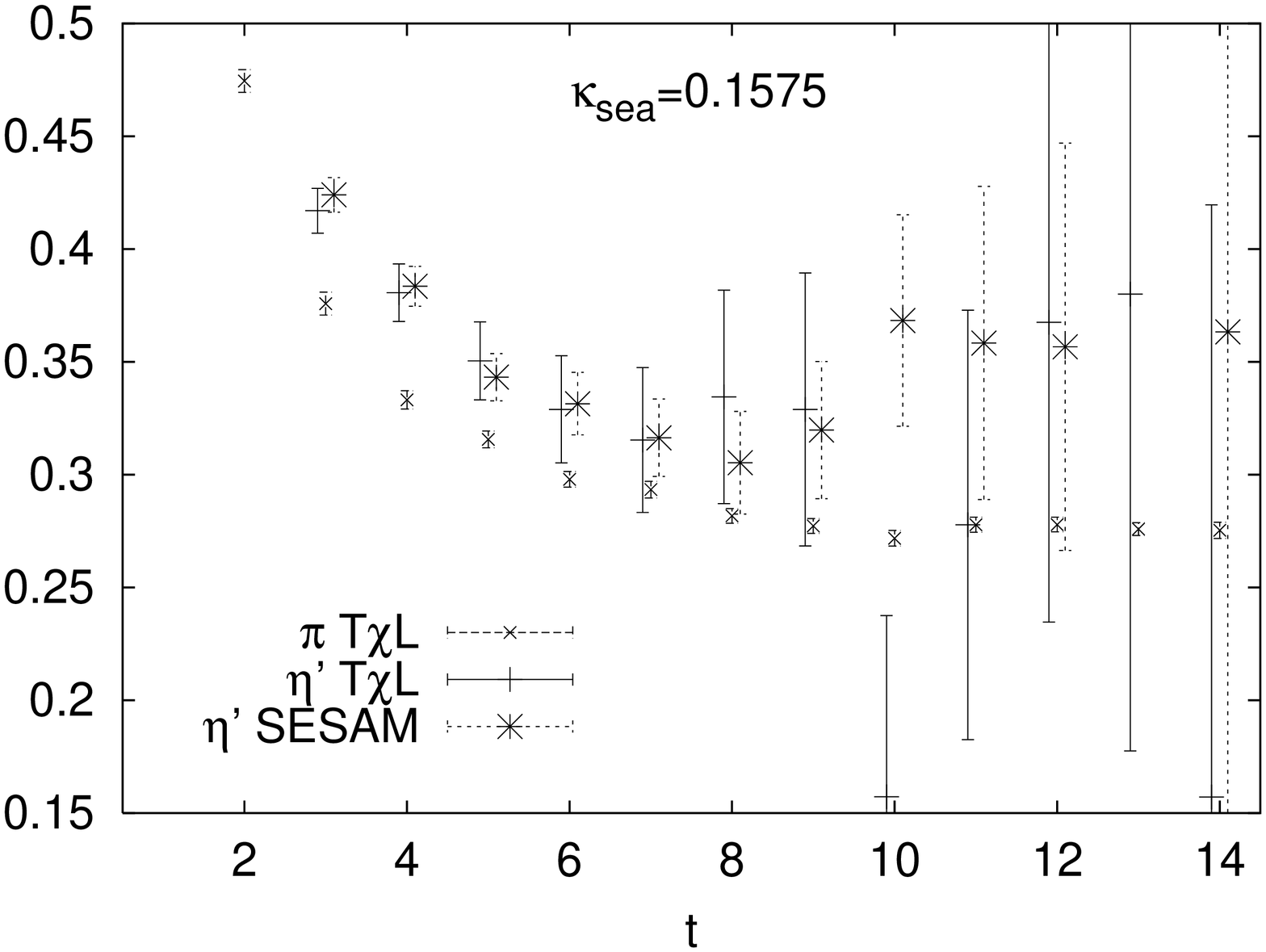,angle=270,width=\columnwidth}}
\centering{\epsfig{figure=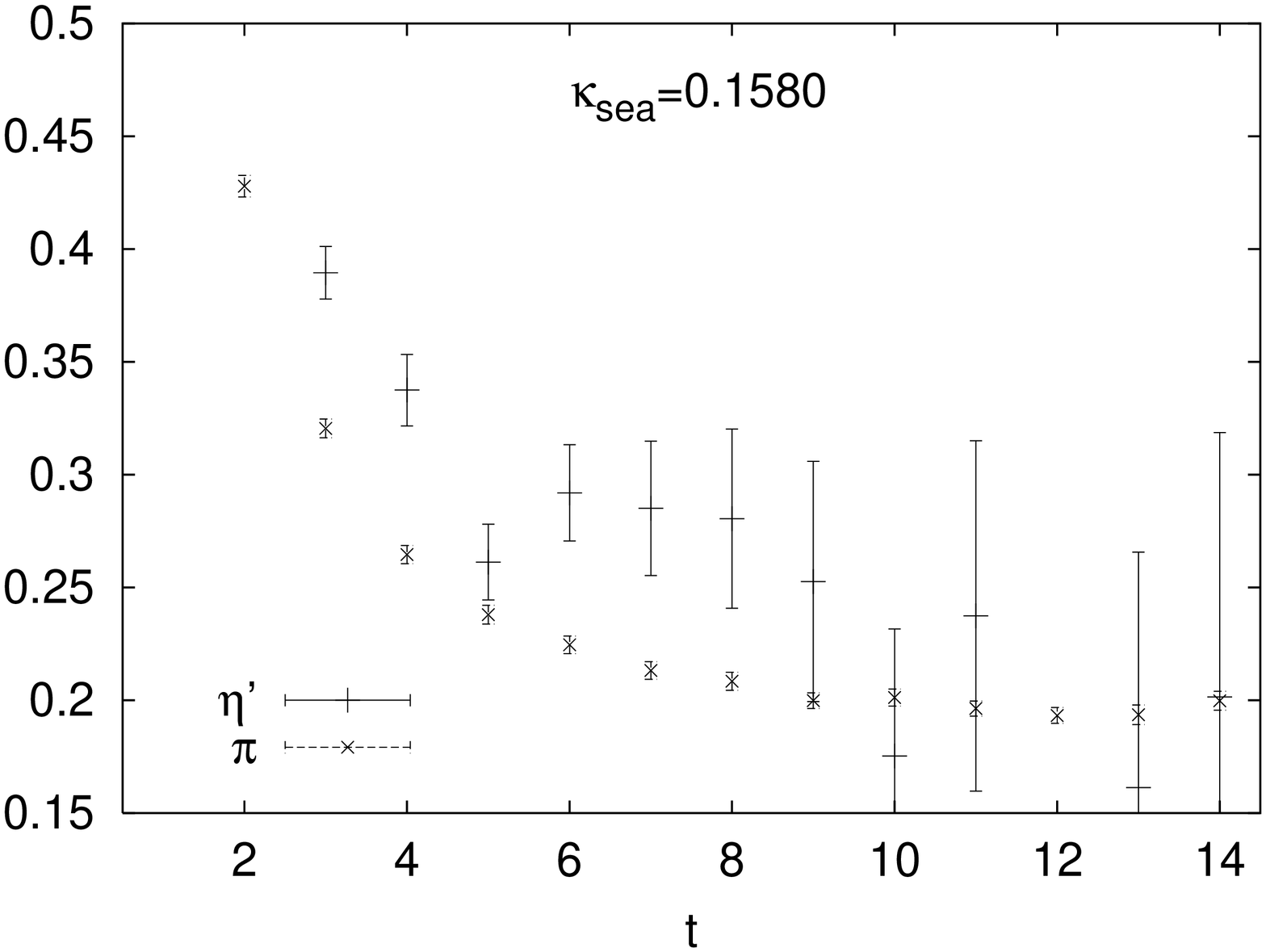,angle=270,width=\columnwidth}}
\caption{Plateau formation in the effective $\eta'$- and $\pi$ masses with
  smeared operators at various   sea quark masses. At $\kappa_{sea} = 0.1575$,
$\eta '$-results from  SESAM and \txl configurations are plotted separately.}
\label{fig:eta_pi_sm}
\end{figure}
Let us now scrutinize the situation by turning to the analysis of effective
masses as extracted from Eq.~(\ref{eq:effmasses}). We have seen in
ref.~\cite{osaka} that, given our sample sizes, the use of local sources and
sinks does not provide sufficient resolution to reveal such plateau formation
in the effective flavour singlet mass plots.  This situation is illustrated in
Fig.~\ref{fig:eff_loc_smear} where we confront, at a particular intermediate
sea quark mass ($\kappa_{sea} = 0.1570$), effective pseudoscalar masses
obtained both with and without   operator smearing.  Obviously, with
pointlike operators, one has to resort to {\it bona fide} single $\cosh$
fits on the correlators without any kind of systematic error control on the
extracted flavour singlet `masses'. After source and sink smearing, however,
our data begins to reveal plateau formation in the singlet channel.

In Fig.~\ref{fig:eta_pi_sm} we display the evidence for plateau formation
through operator smearing for the remaining sea quark masses in the range
$0.1565 \leq \kappa_{sea}\leq 0.1580$.  Here again the `octet' channel masses
are included for reference in order  to enable judgement on the sensitivity for
mass gap determinations.  We emphasize again that all singlet data are
obtained after symmetric source {\it and} sink smearing, as described above.
It appears that smearing meets the expectation by increasing the ground state
overlap: this opens the window of observation for a mass plateau from $t= 5$
onwards where statistical errors are still `tolerable'.
  By comparing the SESAM and \txl data sets at $\kappa_{sea} = 0.1575$
we find no evidence for a volume effect on $m_{\eta'}$\cite{orth}.

\begin{figure}[t]
\centering\epsfig{figure=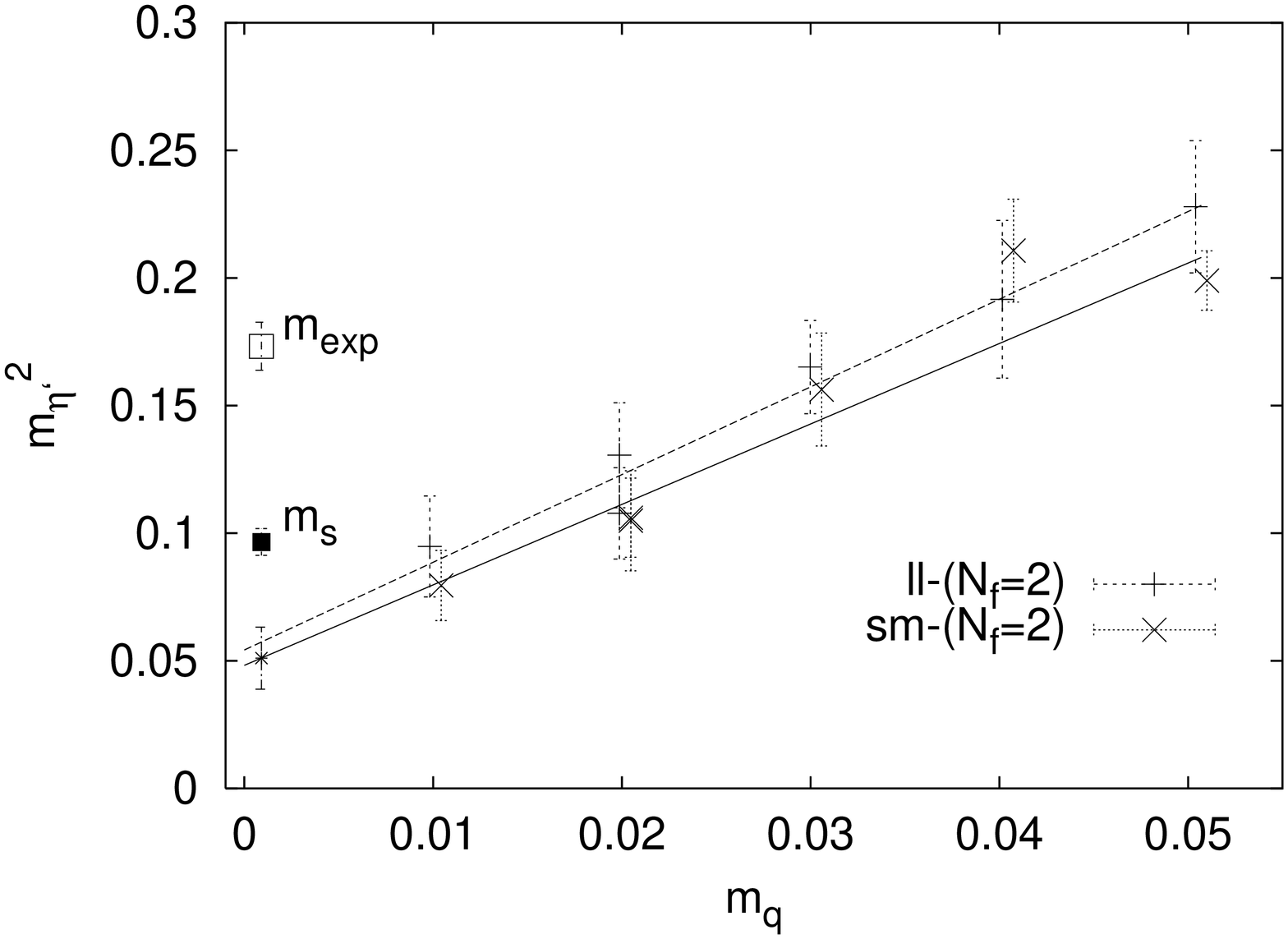,angle=270,width=\columnwidth}
\centering\epsfig{figure=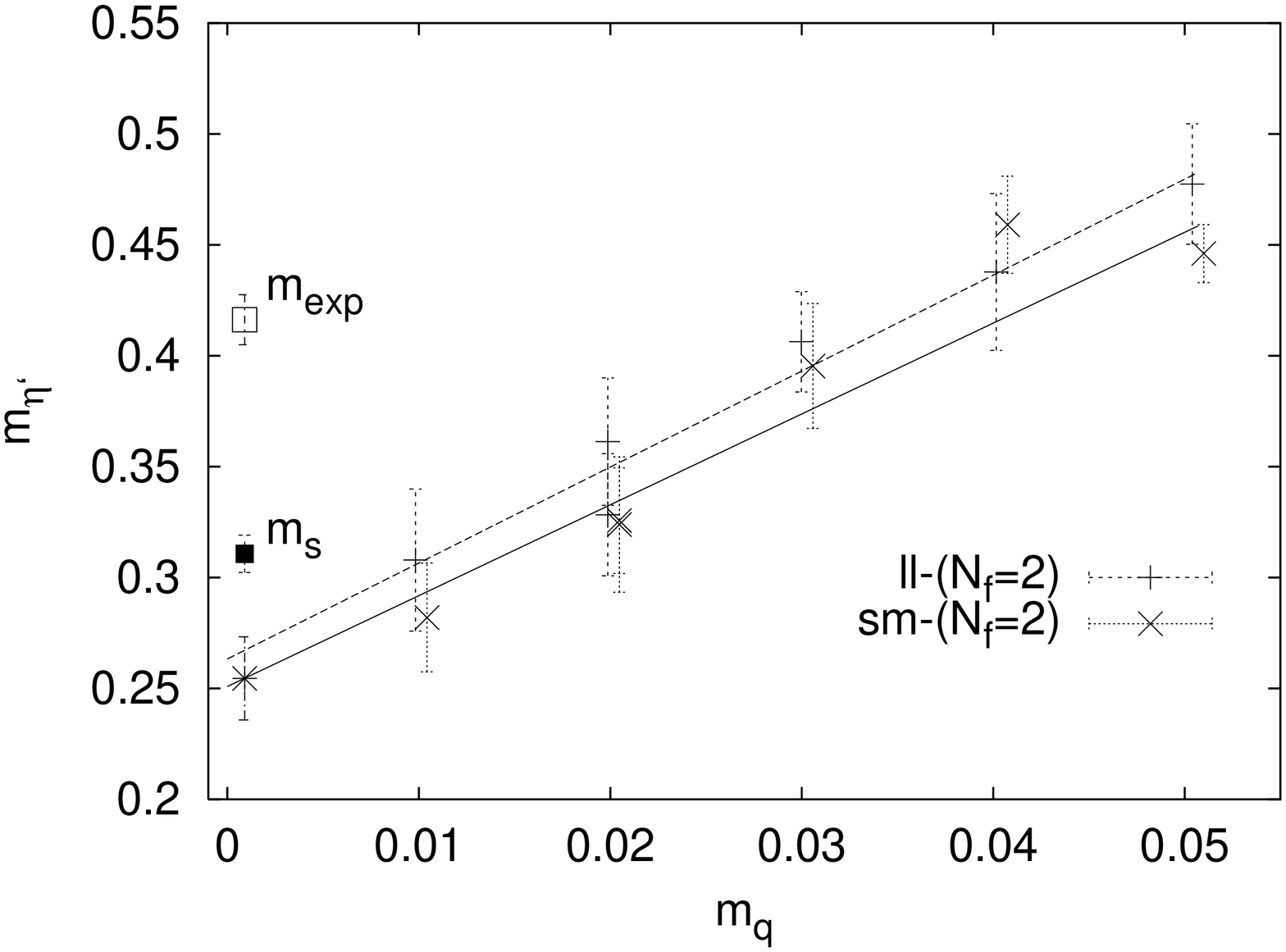,angle=270,width=\columnwidth}
\caption{Chiral extrapolations of $m_{\eta'}^2$ (top figure) and
  $m_{\eta'}$ linear in quark mass.} 
\label{fig:chexeta}
\end{figure}
\section{Physics analysis}
Encouraged by the apparent plateau formation we proceed next to carry out mass
fits based on a {\it single $\cosh$} ansatz, with $t$-ranges as listed in
table~\ref{fittab}.  For reference we have also included information about the
fit-ranges previously used  with local sources~\cite{osaka}.

\subsection{Chiral extrapolations}
\begin{table}[b]
\begin{center}
\caption{$\eta'$ and $m_0$ results \label{tab:mevmasses}} 
\hspace{0.2cm}
\begin{tabular}{|cccccc|}
        \hline
        Ensemble & Fit & $m_{\eta'}$ & $m_0$ & $M_{\eta'}$ [MeV] & $M_0$ [MeV] \\
        \hline
        $N_f=2$ & $m$-ll   & .267(23)  & .251(43) & 615(53) & 576(99) \\
        \hline
        $N_f=2$ & $m^2$-ll & .239(37)  & .245(40) & 551(85) & 565(92) \\
        \hline
        $N_f=2$ & $m$-sm   & .255(19) & .205(43) & 587(44) & 472(99) \\
        \hline
        $N_f=2$ & $m^2$-sm & .226(25)  & .222(26) & 520(58) & 510(62) \\
        \hline
      \end{tabular}
    \end{center}

  \end{table}
Because of the well-known technical limitations of the hybrid Monte Carlo
algorithm~\cite{1999:Kennedy} the SESAM and T$\chi$L configurations
correspond to two mass-degenerate light sea quark flavours ($N_f = 2$), with
the unrenormalized mass value 
\begin{equation}
m_q=M_qa = 1/2(\kappa^{-1} - \kappa_{c}^{-1}) \;.
\end{equation}
From our previous light spectrum analysis~\cite{sesam:masses} we quote the
lattice spacing 
\begin{equation}
a_{\rho}^{-1}(\kappa_{light}) = 2.302(64)\mbox{GeV} 
\end{equation}
and the critical and physical light quark $\kappa$ values:
\begin{equation}
\kappa_c = 0.158507(44)\quad, \quad \kappa_{light}= 0.158462(42) \; .
\end{equation}
Our data do not allow to decide whether it is $m_{\eta'}^2$ or $m_{\eta'}$
that  follows a
linear quark mass dependence: as shown in Fig.~\ref{fig:chexeta}, both
ans\"atze render $\chi^2/d.o.f. \simeq {\cal O}(1)$. We emphasize in this
context that we make {\it no distinction} between sea and valence quarks as we
choose the quark masses in the fermion loops to equal the sea quark masses
(symmetric extrapolation in the sense of ref.~\cite{sesam:masses}).

We display the results on the $\eta '$ mass and the mass gap
\begin{equation}
m_0^2= m_{\eta'}^2-m_{\pi}^2 \;,
\label{eq:gap}
\end{equation}
for both forms of extrapolation in table~\ref{tab:mevmasses}.  For
com\-pa\-rison, we have also included previous estimates as obtained by using
local sources~\cite{osaka}.
\begin{figure}
\centering\epsfig{figure=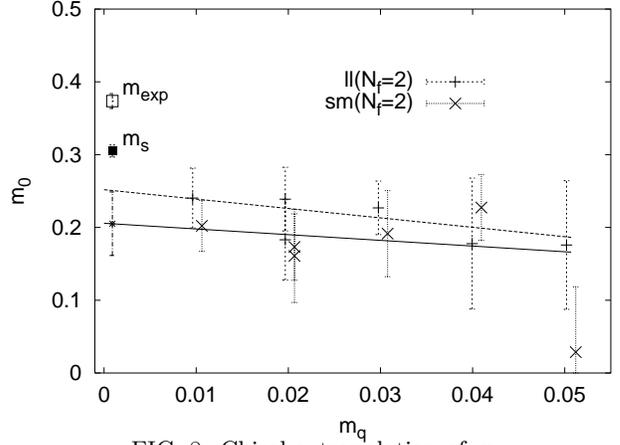,angle=270,width=\columnwidth}
\caption{Chiral extrapolation of $m_0$.}
\label{fig:chexm0}
\end{figure}
\subsection{Comparison to experiment}
In the $N_f = 2$ world of our simulations, according to Eq.~(\ref{eq:greens}),
we would not expect to encounter the full effect of Zweig rule forbidden
diagrams, and hence we anticipate to underestimate the real world $\eta '$
mass, $m_{\exp}$, (plotted as open  squares in Fig.~\ref{fig:chexeta}).

From the experimental mass splitting
\begin{equation}
M_{0,n_F=3}^2 \equiv M_{\eta',n_F=3}^2 - M_{8}^2, \quad M_8^2 \equiv 2M^2_K-M^2_{\eta}\; ,
\label{eq:m0q}
\end{equation}
we therefore compute, in the spirit of the Witten-Veneziano formula,
% eq.~\ref{eq:witten_veneziano},
\begin{equation}
\label{eq:witten_veneziano}
M_0^2 = 2 N_f \chi_q/F_{\pi}^2 \; ,
\end{equation}
 the `pseudoexperimental' value, $M_s$, in our $N_f = 2$ world:
\begin{equation}
M_{s}^2 = 2/3\,M_{0,n_F=3}^2 + M_{\pi}^2 = (715 \; \mbox{MeV})^2 \; .
\label{eq:reduced}
\end{equation}
\begin{figure}[b]
\centering\epsfig{figure=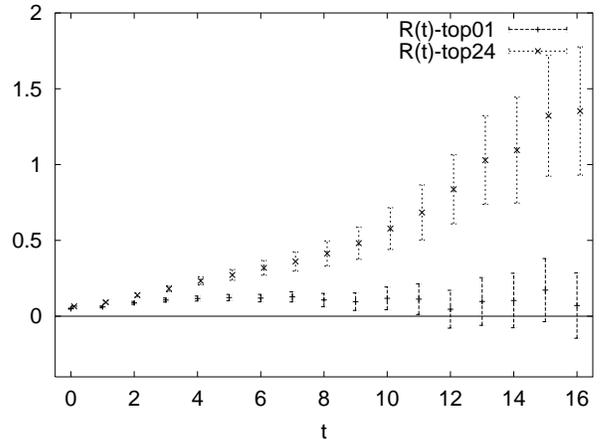,angle=270,width=\columnwidth}
\caption{Ratio of $C_{disc}/C_{conn}$ for $\kappa_{sea} = .1575$ (top figure)
% and effective $\eta'$  (middle) and $\pi$ masses 
with cuts in  topological charge as explained in the
text. }
\label{fig:topratio}
%\end{center}
\end{figure}
\begin{figure}[t]
%\centering\epsfig{figure=./PLOTS/ratiols_k1575_topa0124.eps,width=\columnwidth,angle=270}
\centering\epsfig{figure=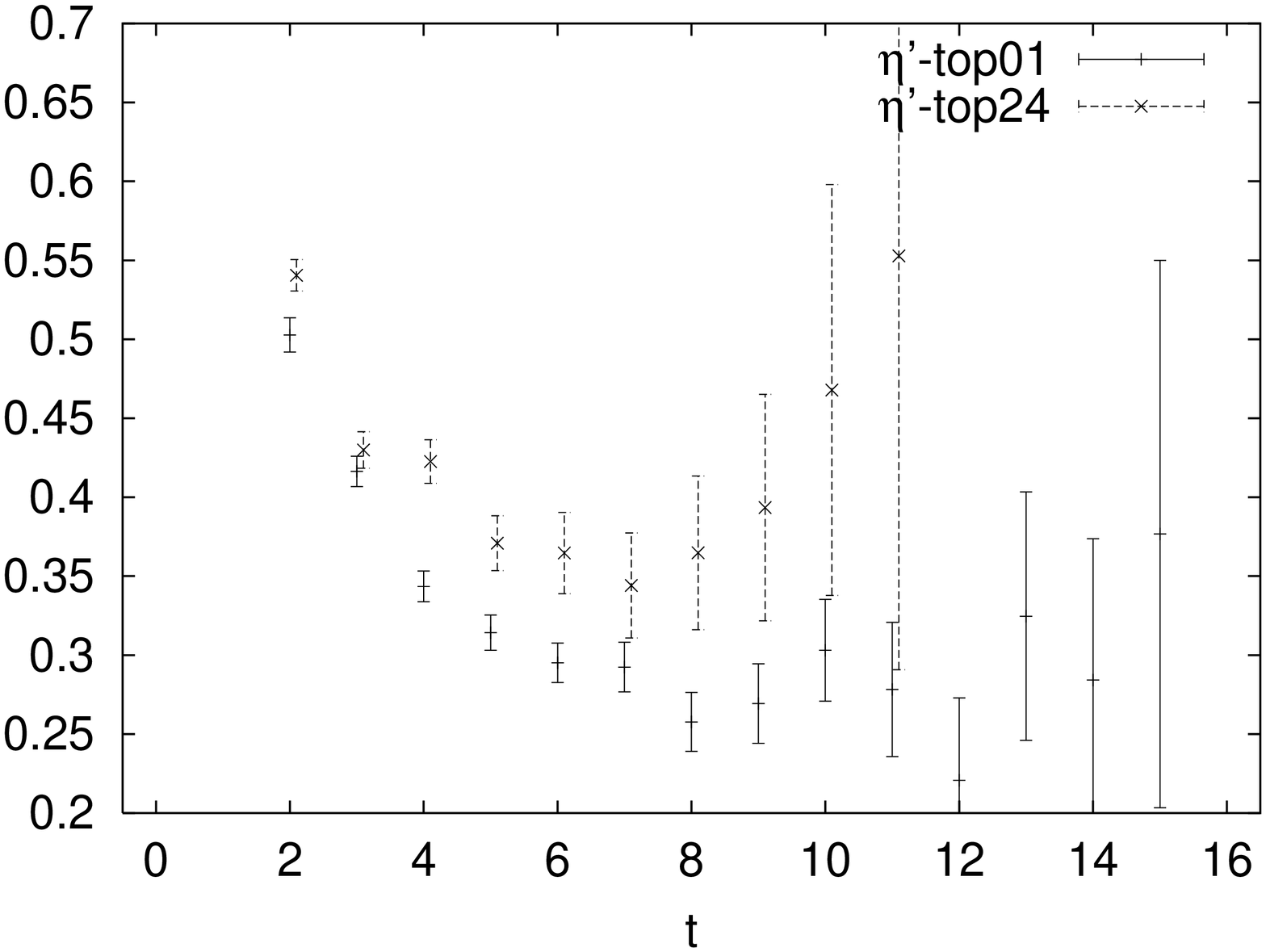,angle=270,width=\columnwidth}
\centering\epsfig{figure=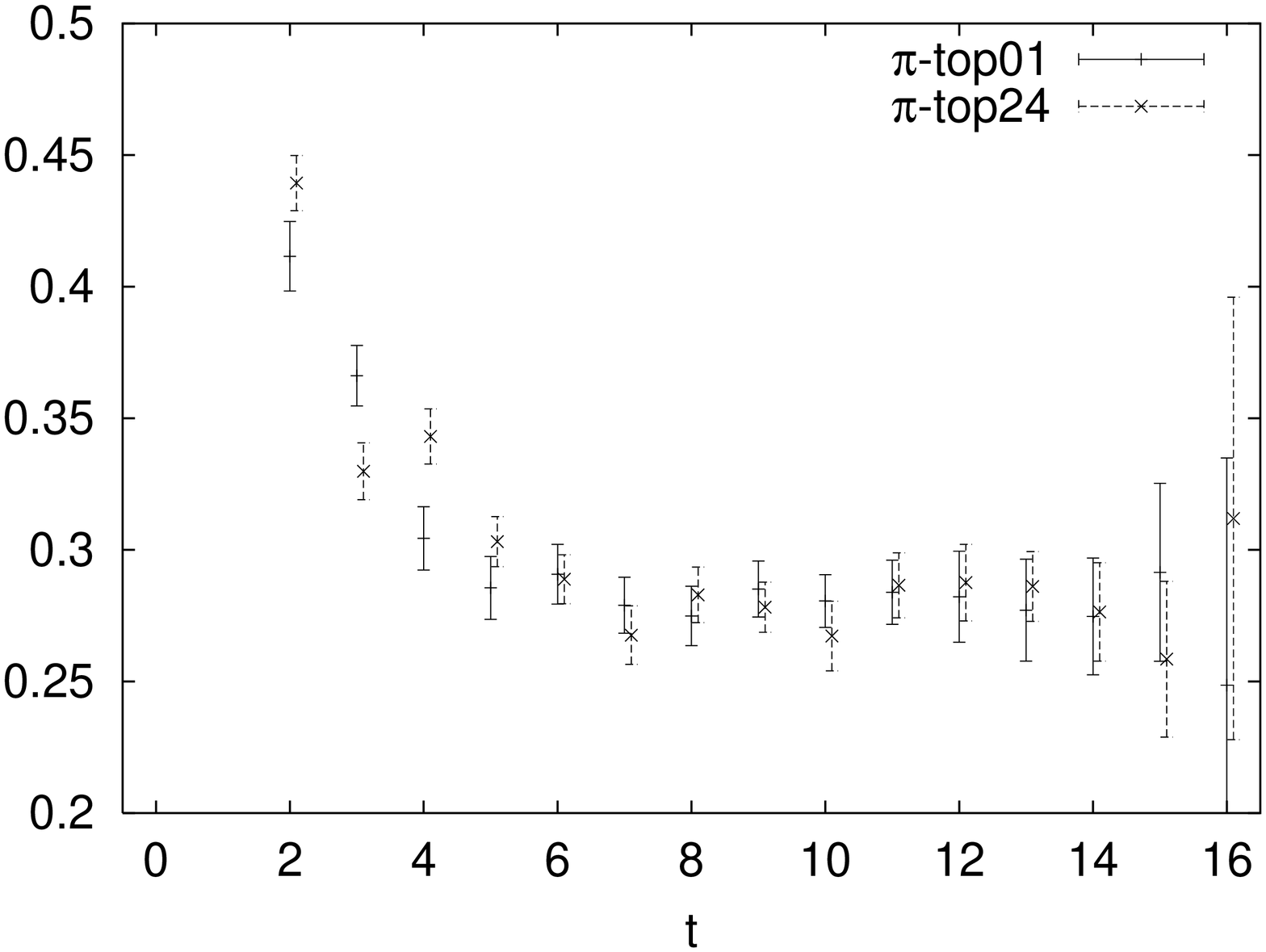,angle=270,width=\columnwidth}
\caption{
%Ratio of $C_{disc}/C_{conn}$ (top figure)
Effective $\eta'$  (top) and $\pi$ (bottom)  masses 
for $\kappa_{sea} = .1575$,
with cuts in  topological charge as explained in the
text.}
\label{fig:topmasses}
%\end{center}
\end{figure}
This value corresponds in lattice units to the full squares marked `$m_s$' in
the two alternative chiral extrapolations shown in Fig.~\ref{fig:chexeta}.
Let us compare this latter value with the lattice $N_f = 2$-prediction in
terms of numbers: when we set the scale by the $\rho$ mass, the extrapolation
of $m_{\eta'}^2$ to the physical quark mass yields the value
\begin{equation}
M_{\eta'}^2 = ( 520^{+125}_{-58} \mbox{MeV})^2
\label{eq:etamass}
\end{equation}
at our lattice spacing.  We have linearly added the difference between the two
extrapolations (of $m_{\eta'}$ and $m_{\eta'}^2$) to the statistical error to
accomodate systematic uncertainties.  The difference between this value and
the 715~MeV of Eq.~(\ref{eq:reduced}) may  be attributed to finite-$a$
effects.  For comparison, we also performed a linear extrapolation of the mass
splitting, $m_0$, as given in Eq.~(\ref{eq:gap}) which shows only little quark
mass dependency (see Fig.~\ref{fig:chexm0}).
This is consistent with the weak dependence of $F_{\pi}$ on
the quark mass observed on SESAM configurations~\cite{sesam:masses}.
 From the value $F_{\pi}=116(8)$~MeV
obtained for $N_F=2$ QCD at our lattice spacing and
Eq.~(\ref{eq:etamass})  we obtain the estimate
$\chi_q=(170_{-15}^{+27}\,\mbox{MeV})^4$ for the quenched
topological susceptibility of Eq.~(\ref{eq:witten_veneziano}), which is 
consistent with the value 180~MeV from 3 flavour
phenomenology~\cite{witten_veneziano}.

\subsection{Impact of topology}

The Witten-Veneziano mass formula, Eq.~(\ref{eq:witten_veneziano}), relates the
difference between the $\eta'$ mass and the flavour non-singlet pseudoscalar
mass to the topological susceptibility $\chi_q$ of the {\it quenched} gauge
vacuum.  This scenario motivates us to investigate in {\it full} QCD whether
the ratio $ R_Q(t)= C_{disc}(t)_Q/C_{conn}(t)_Q$, whose deviations from zero
give rise to the observed mass gap, is correlated with $|Q|$, the modulus of
the topological charge, configuration by configuration.  If we restricted our
analysis for instance to gauge configurations with $Q=0$ only, the topological
susceptibility, $\chi=\langle Q^2\rangle/V$, determined on this subensemble,
would vanish as well and we might expect the $\pi$ to be mass degenerate with
the $\eta'$. On the other hand, if we rejected configurations with small
$|Q|$-values, the effective $\chi$ on the remaining sample would be enhanced
and the generated mass gap increased.

In Fig.~\ref{fig:topratio} we show, for $\kappa_{sea} = 0.1575 $ and the small
lattice, the quantity $R_Q(t)$ with cuts applied according to $|Q| \leq 1.5$
(top01) and $|Q| > 1.5$ (top24), the topological charge being determined as in
ref.~\cite{tunneling}. The value of $1.5$ was chosen such as to obtain two
ensembles {\it of comparable statistics} in topologically different vacuum
sectors.  We do find a definite dependence of $R_Q$ on $|Q|$.  Note in
particular that the disconnected piece vanishes in the vacuum sector with
small values of $|Q|$!  This feature reflects itself of course in the
corresponding effective masses of the flavour singlet and non-singlet mesons.
This is shown in Fig.~\ref{fig:topmasses}: the restricted flavour singlet
mass, $m_{\eta'}|_{|Q| \leq 1.5}$, turns out to be identical to the `octet'
meson mass $m_3 = m_{\pi}$. Accordingly, the flavour singlet-non-singlet mass
gap is due to nontrivial topological vacuum structures.

On the other hand, the `octet' meson mass appears to be not at all sensitive
to such restrictions to  topological sectors; this seems to be a general
feature of flavour non-singlet light hadron spectrum observables
~\cite{toappear}.
\section{Summary and conclusions}
Using smeared operators and reasonable source statistics on the SESAM and \txl
samples of QCD vacuum configurations we found clear indications of  plateau
formation in the effective flavour singlet pseudoscalar mass plot in the
intermediate $t$-regime.  The $\eta'$ mass is definitely sensitive to the
topological structure of the QCD vacuum. In our two-flavour simulation its
actual value after chiral extrapolation turns out to be in qualitative
agreement with the expectation from experiment, but further studies are needed
to pin down finite-$a$  effects.

At this stage the statistical errors on the singlet masses are mostly due to
gauge field fluctations and  by a factor $\simeq 5$ larger than for the
non-singlet ones.  Clearly, the next generation teracomputers will open the
door to go for lattice determinations of Zweig-rule forbidden objects
with an accuracy known so far only from light non-singlet hadron spectroscopy.

In the meantime we are working on computational techniques to determine quark
loops in the regime of quark masses lighter than attained so far~\cite{neff}.

\acknowledgements We enjoyed interesting discussions with John Negele.  TS
thanks Thorsten Feldmann for fruitful conversations.  BO, TS, and WS
appreciate support from the DFG Graduiertenkolleg ``Feldtheoretische und
Numerische Methoden in der Statistischen und Elementarteilchenphysik''.  Our
European collaboration was funded by the EU network ``Hadron Phenomenology
from Lattice QCD'' (HPRN-CT-2000-00145).  GB has been supported by EU grant
HPMF-CT-1999-00353.  The HMC productions were run on APE100 systems at INFN
Roma and at NIC Zeuthen.  We are grateful to our colleagues F. Rapuano and G.
Martinelli for the fruitful \txl -collaboration.  Analysis was performed on
APE100 and CRAY T3E systems at NIC and the University of Bielefeld.

\end{document}